\documentclass[pra,aps,reprint]{revtex4-1}

\usepackage[utf8]{inputenc}
\usepackage{amsmath,amssymb,bm}
\usepackage{graphicx}
\graphicspath{{./img/}}

\newcommand{\sect}[1]{\par\vspace{3ex}\textit{#1}.---\ignorespaces}

\begin{document}

\title{Dispersive readout of adiabatic phases}

\author{Sigmund Kohler}
\affiliation{Instituto de Ciencias de Materiales de Madrid, CSIC, E-28049, Spain}
\date{\today}

\begin{abstract}
We propose a protocol for the measurement of adiabatic phases of periodically
driven quantum systems coupled to an open cavity that enables dispersive
readout.  It turns out that the cavity transmission exhibits peaks at
frequencies determined by a resonance condition that involves the dynamical
and the geometric phase.  Since these phases scale differently with the
driving frequency, one can determine them by fitting the peak positions to
the theoretically expected behavior.  For the derivation of the resonance
condition and for a numerical study, we develop a Floquet theory for the
dispersive readout of ac-driven quantum systems.
The feasibility is demonstrated for two test cases that generalize
Landau-Zener-St\"uckelberg-Majorana interference to two-parameter driving.
\end{abstract}

\maketitle

%
A state vector that undergoes a cyclic evolution in Hilbert space acquires
a phase factor that can be divided into a dynamical and a geometric part
\cite{Aharonov1987a}.  The latter is gauge invariant and in the adiabatic
limit agrees with the phase discovered earlier by Berry for cyclic
adiabatic following \cite{Berry1984a}.  Ever since these phases have
attracted much attention owing to their fundamental significance as well as
for the their usefulness for, e.g., computing electronic properties of
solids \cite{Xiao2010a}.  The Floquet states of an ac driven system are
periodic in time and describe a cyclic solution of
the time-dependent Schr\"odinger equation \cite{Shirley1965a, Sambe1973a}.
Moreover, they can be characterized by a mean energy which is essentially
the dynamical phase, while the geometric phase corresponds to the
difference between the quasienergy and the mean energy \cite{Moore1990a,
Grifoni1998a}.

The direct observation of geometric phases is hindered by two obstacles.
First, phases are commonly visible in interference effects that depend on
the \textit{relative} phase of a superposition.  Second, interference
is sensitive to the total phase without distinguishing between a dynamical
and a geometric contribution.  Therefore, measuring geometric
phases requires the construction of a Hamiltonian for which the dynamical
phase vanishes \cite{Unanyan2004a}.  This can be accomplished also dynamically
by spin-echo ideas by which a $\pi$-pulse inverts the Bloch vector
after half a driving period \cite{Falci2000a, Peng2006a, Leek2007a}.
The present work proposes a measurement scheme for geometric phases which
differs from previous ones in the way the dynamical and the geometric phase
are disentangled.

Landau-Zener-St\"uckelberg-Majorana (LZSM) interference in solid-state
qubits \cite{Shevchenko2010a} is achieved by driving a qubit such that its
excitation probability as a function of the detuning and the driving
amplitude exhibits a characteristic pattern.  The common one-parameter
driving, however, restricts the adiabatic phases to multiples of $2\pi$,
despite that beyond the adiabatic limit non-trivial geometric phases may
emerge.  Since LZSM interferometry became an established
experimental technique, it is natural to use it as starting point and to
extend it by a second driving that enables non-trivial adiabatic phases.

To gain information about a qubit, one frequently employs dispersive
readout based on the coupling to a cavity.  Then the cavity experiences a
frequency shift which depends on the qubit state and can be probed
via the transmission \cite{Blais2004a}.  Theoretically this can be
seen in the qubit-cavity Hamiltonian after a transformation to the
dispersive frame \cite{Blais2004a, Zueco2009b}.  In a more systematic and
generalizable treatment, one relates the cavity transmission to the
susceptibility of the qubit coupled to it \cite{Petersson2012a,
Burkard2016a}.

Here we generalize the theory of dispersive readout to \textit{ac driven}
quantum systems.  We will find that the cavity transmission as a function
of the driving frequency exhibits regularly spaced peaks.  Their distances
relate to the Berry phases of the adiabatic eigenstates.

\sect{Dispersive readout of a driven system}
We consider a driven quantum system, henceforth ``qubit'', with the
$T$-periodic Hamiltonian $H_\text{qb}(t)$ and the driving frequency
$\Omega=2\pi/T$.  Decoherence is taken into account by employing a quantum
master equation that starts from the qubit-bath Hamiltonian
$H_\text{qb-bath} =X\sum_\nu \lambda_\nu(b_\nu^\dagger+b_\nu)$, with
$b_\nu$ describing the modes of a bosonic environment \cite{Leggett1987a,
Hanggi1990a}, for details see
Appendix~\ref{app:susceptibility}.
Dispersive readout \cite{Blais2004a} is enabled by coupling the qubit to an
open cavity such that the central Hamiltonian reads (in units with
$\hbar=1$)
\begin{equation}
\label{Hqbcav}
H = H_\text{qb}(t) + gZ(a^\dagger+a) + \omega_0 a^\dagger a
\end{equation}
with a qubit operator $Z$ and $a$ the usual bosonic
annihilation operator of the cavity mode.  The cavity
couples at both ends $\nu=1,2$ to incoming and outgoing modes.  Starting
again from a system-bath model, one can employ input-output theory
\cite{Collett1984a, Gardiner2004a, Clerk2010a} to obtain the quantum
Langevin equation
\begin{equation}
\label{dota}
\dot a_t = -i\omega_0 a_t -\frac{\kappa}{2} a_t -\sum_{\nu=1,2} \sqrt{\kappa_\nu}
a_{\text{in},\nu} -igZ_t
\end{equation}
with the input fields $a_{\text{in},\nu}$ and the cavity loss rate $\kappa
= \kappa_1+\kappa_2$.  The corresponding time-reversed equation provides
the input-output relation $a_{\text{in},\nu} +a_{\text{out},\nu} =
\sqrt{\kappa_\nu}a$ and the cavity transmission which contains
information about the qubit.

In turn, the qubit experiences a force from the cavity which can be derived
(see Appendix~\ref{app:lrt})
from linear response theory to read
\begin{equation}
\label{lrt}
\langle Z_t\rangle
= g\int dt'\, \chi(t,t') \langle a^\dagger_{t'} + a_{t'}\rangle \,.
\end{equation}
The susceptibility $\chi(t,t') = -i\langle[Z(t),Z(t')]\rangle^{(0)}
\theta(t-t')$ has to be evaluated at time $t'$ in the absence of the
cavity.  Generally for time-dependent systems, such expressions depend
explicitly on both times, while for periodic driving, $\chi(t,t') =
\chi(t+T,t'+T)$.  Consequently, upon introducing the time
difference $\tau=t-t'$ we find that $\chi(t,t-\tau)$ is $T$-periodic in $t$
\cite{Kohler2005a} and, thus, can be written as
\begin{equation}
\label{chikw}
\chi(t,t-\tau) = \sum_k\int \frac{d\omega}{2\pi}\,
e^{-ik\Omega t-i\omega\tau} \chi^{(k)}(\omega) \,.
\end{equation}
This implies for Eq.~\eqref{lrt} the Fourier representation $Z_\omega = g
\sum_k \chi^{(k)}(\omega-k\Omega)(a_{\omega-k\Omega}
+a^\dagger_{\omega-k\Omega})$ which reflects the frequency mixing inherent
in the linear response of the driven quantum system.

Since $a_\omega$ has its dominating contribution at the bare cavity
frequency $\omega_0$, the qubit response $Z_\omega$ mainly contains the
frequencies $\pm\omega_0 + k\Omega$, where contributions with $-\omega_0$
stem from the creation operator $a^\dagger_\omega$.  For the backaction of
the qubit to the oscillator, the component with frequency $\omega_0$
represents the only resonant excitation.  In the good cavity limit
$\kappa\ll\omega_0,\Omega$, we can neglect within a rotating-wave
approximation all non-resonant components, which means that the relevant
qubit response is given by $Z_\omega = g\chi^{(0)}(\omega) a_\omega$.
Inserting this result into Eq.~\eqref{dota}, we obtain
via Fourier transformation an expression for $a_\omega$.  Together with
the input-output relation follows the cavity transmission amplitude
\begin{equation}
\label{tc}
t_c = \frac{\langle a_{\text{out},2}\rangle}{\langle a_{\text{in},1}\rangle}
= \frac{i\sqrt{\kappa_1\kappa_2}}{\omega_0-\omega+g^2\chi^{(0)}(\omega)-i\kappa/2} .
\end{equation}
Its dependence on $\chi$ allows one to acquire information about the qubit
by probing the transmission $|t_c|^2$ \cite{Blais2004a} (for graphical
reasons, the reflection $R=1-|t_c|^2$ will be plotted).

\sect{Floquet theory}
The remaining task is the computation of $\chi^{(0)}(\omega)$. To this end,
we employ the Floquet-Markov formalism developed in
Ref.~\cite{Kohler1997a}.  It starts by diagonalizing
$H_\text{qb}(t)-i\partial_t$ in the Hilbert space extended by the space of
$T$-periodic functions \cite{Shirley1965a, Sambe1973a} to obtain the
Floquet states $|\phi_\alpha(t)\rangle = |\phi_\alpha(t+T)\rangle$, the
quasienergies $\epsilon_\alpha$, the mean energies $\bar E_\alpha$, and the
stationary solutions of the Schr\"odinger equation, $|\psi_\alpha(t)\rangle
= e^{-i\epsilon_\alpha t}|\phi_\alpha(t)\rangle$.  The corresponding
expression for the propagator, $U(t,t') = \sum_\alpha
e^{-i\epsilon_\alpha(t-t')} |\phi_\alpha(t)\rangle
\langle\phi_\alpha(t')|$, allows us to deal with the interaction picture
operators in $\chi(t,t')$.
Moreover, the Floquet states provide a convenient basis for the
Bloch-Redfield master equation \cite{Redfield1957a, Blum1996a} for the
qubit density operator which in this representation eventually becomes
diagonal \cite{Kohler1997a}, $\rho_\text{qb} = \sum
p_\alpha|\phi_\alpha(t)\rangle \langle\phi_\alpha(t)|$.
In contrast to a static system, the probabilities $p_\alpha$ at long times
are not simple Boltzmann factors.

With these ingredients, we find from Eq.~\eqref{chikw} the susceptibility,
\begin{equation}
\label{chi0w}
\chi^{(0)}(\omega) = \sum_{\beta,\alpha,k}
\frac{(p_\alpha-p_\beta)|Z_{\beta\alpha, k}|^2}
{\omega+\epsilon_\alpha-\epsilon_\beta+k\Omega+i\gamma/2},
\end{equation}
where $Z_{\beta\alpha,k}$ denotes the $k$th Fourier component of the
$T$-periodic transition matrix element $Z_{\beta\alpha}(t) =
\langle\phi_\beta(t)|Z|\phi_\alpha(t)\rangle$.
To regularize the Fourier integrals, we have introduced the phenomenological
level broadening $\gamma$ of the Floquet states.

Equation \eqref{tc} together with Eq.~\eqref{chi0w} represents a
generalization of dispersive readout to quantum systems under strong ac
driving (in addition to the weak driving entailed by $a_{\text{in},1}$ via
the cavity).  For details of the derivation, see
Appendix~\ref{app:susceptibility}.

\sect{Resonance condition, geometric phase, and adiabatic limit}
Upon resonant driving of the cavity with frequency $\omega=\omega_0$, the
reflection $R$ assumes its maximum when the real part of the susceptibility
\eqref{chi0w} vanishes.  This is the case when the quasienergies and the
driving frequency obey the condition
\begin{equation}
\label{reso1}
\epsilon_\beta - \epsilon_\alpha = \omega_0 + k\Omega_k ,
\end{equation}
where $\Omega_k$ is the $k$th resonance (notice that smaller $\Omega_k$
have larger index).

The second cornerstone of our protocol stems from the relation between
Floquet states and geometric phases. The former are time-periodic and
during each cycle they acquire the phase $\epsilon_\alpha T = \bar E_\alpha T
-\gamma_\alpha$, where $\gamma_\alpha = \int_0^T dt \langle
\phi_\alpha(t)|i\partial_t|\phi_\alpha(t)\rangle$ is the (non-adiabatic)
geometric phase of the Floquet state $|\phi_\alpha(t)\rangle$
\cite{Aharonov1987a, Moore1990a, Grifoni1998a}.
To avoid difficulties with the quasienergies for small frequencies
\cite{Hone1997a}, we substitute the $\epsilon_\alpha$ by $\bar E_\alpha
-\gamma_\alpha/T$, where both $\bar E_\alpha$ and $\gamma_\alpha$ have a
well-defined adiabatic limit in which $\gamma_\alpha$ becomes the Berry
phase \cite{Aharonov1987a, Moore1990a}.  Then the resonance condition
\eqref{reso1} becomes
\begin{equation}
\frac{1}{\Omega_k}
= C_{\beta\alpha}\Big( k +\frac{\gamma_{\beta\alpha}}{2\pi}\Big)
\label{reso2}
\end{equation}
with $C_{\beta\alpha}=(\bar E_\beta-\bar E_\alpha - \omega_0)^{-1}$ and
the geometric phase $\gamma_{\beta\alpha} \equiv \gamma_\beta-\gamma_\alpha$.

This result suggests for the dispersive readout of adiabatic phases the
following strategy.  One considers low frequencies, such that the terms on
the right-hand side of Eq.~\eqref{reso2} assume their adiabatic limit
(possible corrections are of the order $\Omega$).  Then the cavity
reflection $R(\Omega)$ exhibits peaks whose positions as a function of
their index $k$ can be fitted to the expected linear behavior.  This
provides the adiabatic phase $\gamma_{\beta\alpha}$, the coefficient
$C_{\beta\alpha}$ and, thus, the dynamical phase $-(\bar E_\beta-\bar
E_\alpha)T$.  The index $k$ of the probed resonances may contain an offset
$k_0$ which is irrelevant, because it changes $\gamma_{\beta\alpha}$ merely
by an irrelevant multiple of $2\pi$.

In the low-frequency limit, the system will eventually reside in the
Floquet state with the smallest mean energy, labelled with
$\alpha=0$.  Then dispersive readout probes the geometric phases
$\gamma_{\beta0} = \gamma_\beta-\gamma_0$.

\sect{Two-level system as test case}
The paradigmatic example for a Berry phase is the one of a two-level system
with the pseudo-spin Hamiltonian $H_\text{qb} =\frac{1}{2} \mathbf{B}(t)
\cdot\bm{\sigma}$ and the periodically time-dependent ``magnetic field''
$\mathbf{B}(t+T) = \mathbf{B}(t)$.  As is well known \cite{Sakurai}, the
adiabatic ground state of $H_\text{qb}$ acquires a geometric phase that
equals the solid angle of the curve $\mathbf{B}(t)$ with respect to the
origin.  This implies that for a smooth behavior of the adiabatic phase as
a function of a control parameter, $H_\text{qb}(t)$ must contain all three
Pauli matrices.  Let us therefore consider the qubit Hamiltonian
\begin{equation}
\label{Hqb}
H_\text{qb}(t) = \frac{\Delta}{2}\sigma_x + \frac{\epsilon}{2}\sigma_z
+A\sigma_z\cos(\Omega t) + B\sigma_y\sin(\Omega t)
\end{equation}
with tunnel matrix element $\Delta$, detuning $\epsilon$, and driving
amplitudes $A$ and $B$.  For $B=0$ this model has been widely studied in
the context of LZSM interference \cite{Shevchenko2010a}.
To complete the model, we choose for the qubit-cavity
coupling the operator $Z=\sigma_z$, while the qubit-bath interaction is
specified by $X=\sigma_x$.

\begin{figure}
\centerline{\includegraphics{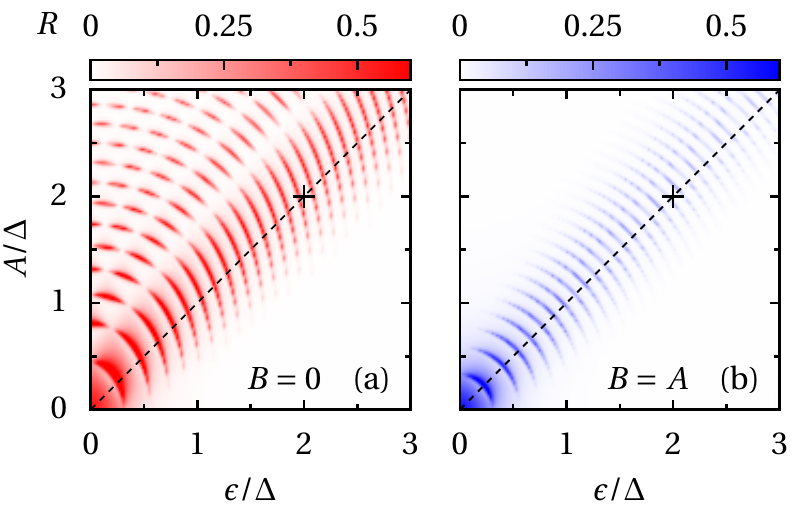}}
\caption{Reflection of the resonantly excited cavity as a function of the
detuning and the amplitude for one-parameter driving (a) and
two-parameter driving (b) of the qubit with frequency $\Omega=0.1\Delta$.
Cavity frequency and line width are $\omega_0 = 0.95\Delta$ and
$\kappa=\omega_0/1000$, while the coupling constants to the
qubit and the bath read $g=0.005\Delta$ and $\alpha=0.01$.
Crosses and dashed lines mark the values used in Figs.~\ref{fig:peaks} and
\ref{fig:berryphase}(a), respectively.
}
\label{fig:patterns}
\end{figure}

For later comparison, we start with one-parameter driving ($B=0$) for
which $\gamma_{10}$ vanishes.  Figure~\ref{fig:patterns}(a) shows the
resulting cavity reflection as a function of the detuning and the driving
amplitude.  It has the typical shape of a low-frequency LZSM
interference pattern \cite{Shevchenko2010a}.  For two-parameter driving
with equal amplitudes, $B=A$, [Fig.~\ref{fig:patterns}(b)], only the
resonances close to $A\approx |\epsilon|$ are well visible.
Henceforth we concentrate on this region.

\begin{figure}
\centerline{\includegraphics{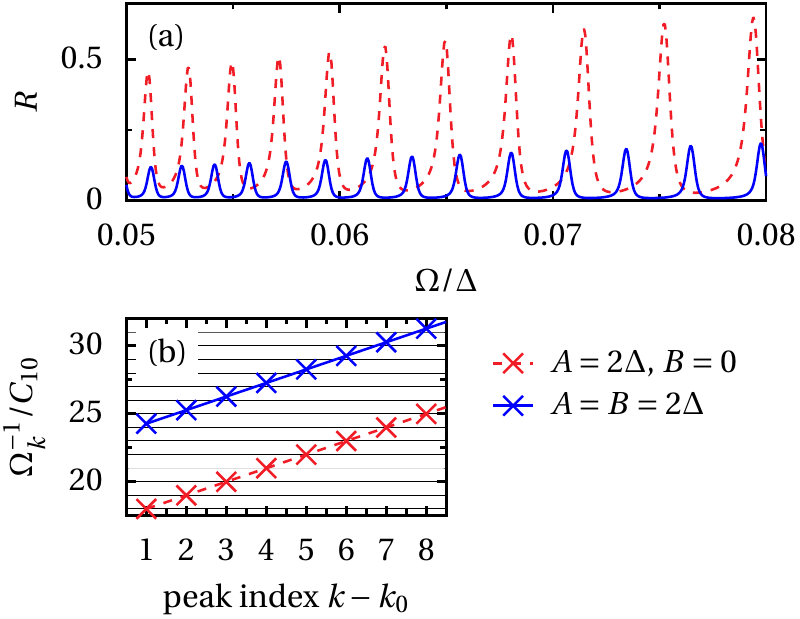}}
\caption{(a) Cavity reflection as a function of the driving frequency
for one-parameter (dashed) and two-parameter driving (solid) with
$A = \epsilon = 2\Delta$.  All other parameters are as in
Fig.~\ref{fig:patterns}.
(b) Corresponding reciprocal peak positions as a function of their index
with an unknown offset $k_0$.
The scaling factor $C_{10}$ stems from fitting the data to
Eq.~\eqref{reso2} with $\gamma_{10}$ as a second fit parameter.  The
vanishing Berry phases for one-parameter driving correspond to integer
values marked by horizontal lines.  Lines connecting the crosses
are a guide to the eye.
}
\label{fig:peaks}
\end{figure}

According to our readout protocol, we consider fixed parameters and
amplitudes and vary the driving frequency.  The resulting cavity reflection
[Fig.~\ref{fig:peaks}(a)] exhibits non-equidistant peaks.  We fit their
positions to the behavior predicted by Eq.~\eqref{reso2} and, in doing so,
determine from the spectrum both $C_{10}$ and the adiabatic
phase $\gamma_{10}$.  The accordingly scaled inverse peak positions are
analyzed in Fig.~\ref{fig:peaks}(b).  For one-parameter driving, they
assume integer values in agreement with the fact that the adiabatic phases
are multiples of $2\pi$.  For two-parameter driving, according to
Eq.~\eqref{reso2}, the values are shifted to non-integer values.

\begin{figure}
\centerline{\includegraphics{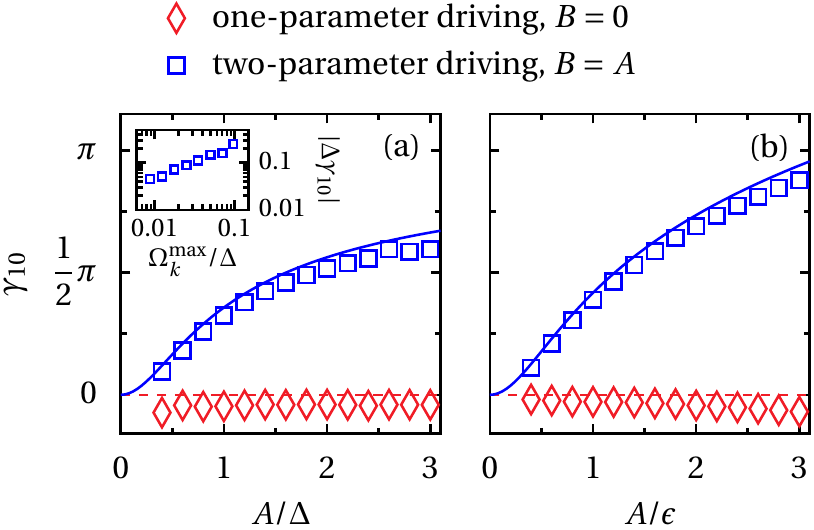}}
\caption{(a) Adiabatic phases as a function of the driving amplitude $A$
for the detuning $\epsilon=A$ marked in Fig.~\ref{fig:patterns} by dashed
lines.  The symbols depict the values reconstructed from the positions of
the first 10 resonance peaks with $\Omega<0.05\Delta$.  The lines
correspond to $\gamma_{10}^\text{ad}$ computed with the adiabatic
eigenstates of $H_\text{qb}(t)$.
Inset: Deviation $|\gamma_{10}-\gamma_{10}^\text{ad}|$ as a function of the
largest $\Omega_k$ used for the fitting at $A=B=\epsilon=2\Delta$.
(b) Corresponding plot for the alternative Hamiltonian
\eqref{HqbB} for constant detuning $\epsilon$ as a function of the
driving amplitude $A$ with the tunneling $\Delta = A/1.2$.
}
\label{fig:berryphase}
\end{figure}

To verify that adiabatic phases can be obtained in a broad parameter range
by the measurement of spectra such as the one in Fig.~\ref{fig:peaks}(a),
we repeat this procedure for the amplitudes and detunings marked in
Fig.~\ref{fig:patterns} by dashed lines.  The symbols in
Fig.~\ref{fig:berryphase}(a) depict the phases reconstructed in this way.
The lines, by contrast, are obtained by diagonalizing
$H_\text{qb}(t)$ and numerically evaluating
$\gamma_\alpha^\text{ad} = \int_0^T dt \langle
u_\alpha(t)|i\partial_t|u_\alpha(t)\rangle$ for each adiabatic eigenstate
$|u_\alpha(t)\rangle$ \cite{on_bargmann, Simon1993}.

Typically the ``measurement'' deviates from the theoretically expected
values by up to $0.03\pi$.  This difference diminishes with smaller driving
frequency
[see inset of Fig.~\ref{fig:berryphase}(a)]
and therefore can be attributed mainly to non-adiabatic corrections.
Moreover, for tiny driving amplitudes, the resonance peaks may not be very
pronounced such that their positions cannot be determined well.
Nevertheless, the numerical simulation of the experiment confirms the main
message of this work, namely that the smooth growth of $\gamma_{10}$ with
the amplitudes $A$ and $B$ can be determined from the resonance peaks in
the dispersive readout, even when the system is not operated in the very deep
adiabatic regime.

It is worthwhile to estimate the precision required in a possible experiment.
In the fitting procedure, one determines $k+\gamma_{10}/2\pi = (\Omega_k
C_{10})^{-1}$ for $k\sim20$ and $0\leq|\gamma_{10}|<2\pi$.  Hence, to identify
$\gamma_{10}$ with a deviation clearly below $2\pi$, the relative error of
the peak positions $\Omega_k$ should not exceed~1\%.  Then the accurracy is
roughly as in the experiment of Ref.~\cite{Leek2007a}.

\sect{Qubit with time-dependent tunnel phase}
As an alternative setup, we consider the Hamiltonian
\begin{equation}
\label{HqbB}
H_\text{qb}'(t) = \frac{\Delta}{2}\sigma_x + \frac{\epsilon}{2}\sigma_z
+A\sigma_x\cos(\Omega t) -B\sigma_y\sin(\Omega t) \,,
\end{equation}
where $\sigma_z$ refers to the charge degree of freedom, while $\sigma_x$
and $\sigma_y$ describe tunneling.  Thus for $A=B$, the driving in
Eq.~\eqref{HqbB} corresponds to a tunnel matrix element
$\propto\exp(i\Omega t)$.  This can be achieved with two quantum dots that
are connected by two independent paths \cite{Gustavsson2007a}.  Then a
time-dependent flux between paths yields the driving in $H_\text{qb}'(t)$.
A further implementation has been proposed \cite{Falci2000a, Peng2006a} and
realized \cite{Leek2007a} by driving a qubit resonantly with a signal that
possesses a linearly growing phase.

Since the driving couples to $\sigma_x$, we now keep the detuning
$\epsilon$ fixed and vary the static tunneling $\Delta$.  The resulting
LZSM patterns are similar to those in Fig.~\ref{fig:patterns}, but with the
main resonance lines for slightly larger amplitudes
(see Appendix \ref{app:patterns}).
The directly computed adiabatic phase and the one extracted from the peak
positions are compared in Fig.~\ref{fig:berryphase}(b).  Their agreement is
similar to the one found for $H_\text{qb}(t)$.

\sect{Conclusions}
We have proposed a protocol for the dispersive readout of adiabatic phases
of time-dependent quantum systems.  Its main difference to previous
proposals \cite{Unanyan2004a, Falci2000a, Peng2006a} and experiments
\cite{Leek2007a} lies in the treatment of the dynamical phase.  While former works
employed spin echo ideas or sophisticated Hamiltonians to physically
eliminate the dynamical phase from the quantum dynamics, the present scheme
provides the values of both phases from an analysis of the resonance
spectrum.  Therefore the protocol is applicable even when a dynamical phase
cannot be avoided or reversed.  The essential ingredients are,
first, that the dynamical and the adiabatic phase as a function of the
driving frequency scale differently and, second, that the resonance
condition for dispersive readout of a driven quantum system can be
expressed in terms of these phases.  The importance of the scaling behavior
relates the scheme to a proposal for measuring Chern numbers by considering
the response of a quantum system to a quench as a function of the velocity
at which the parameters are changed \cite{Gritsev2012a, Schroer2014a}.
There, however, the measurement relies on a more involved quantum state
tomography.  Here, by contrast, the conceptually simpler dispersive readout
signal is sufficient.

We have demonstrated the feasibility of the protocol by its numerical
simulation for realistic solid-state qubits.  The results indicate that
even though the system is not operated in the deep adiabatic regime,
present technology allows recovering the phases with a precision of roughly
$\pi/30$ as follows from a rather conservative estimate.  This may be
improved by employing future qubits with less decoherence which enables
smaller driving frequencies and yields sharper peaks.

For the theoretical description, we have developed a Floquet approach for
the dispersive readout of periodically driven quantum systems.  Its
cornerstone is the identification of the relevant Fourier component of the
qubit susceptibility.  This theory was essential in two respects.  First,
it allowed the computation of the cavity reflection.  Second, it provided
insight to the measurement of driven quantum system as well as the
resonance condition for the peaks in the transmission spectrum.

The recent experimental success with the dispersive readout of qubits in
the context of LZSM interference indicates the feasibility of similar
measurements with a second driving parameter.  In turn, the prospects of
observing in this way not only novel interference patterns, but also a
quantity of fundamental interest such as the Berry phase may motivate
researchers to attempt such experiments.

\begin{acknowledgments}
This work was supported by the Spanish Ministry of Economy and
Competitiveness via Grant No.\ MAT2014-58241-P.
\end{acknowledgments}

\appendix

\section{Response of a strongly driven qubit to an additional weak driving}
\label{app:lrt}

We consider a strongly driven quantum system with the time-dependent
Hamiltonian $H_0(t)$ and a coupling to a heat bath.  Then generally
the reduced density operator $\rho_0(t)$ is time-dependent
as well and may describe a situation far from equilibrium.
In addition, the system is weakly driven by a force $f(t)$ entering via
an operator $Y$, such that the Hamiltonian becomes $H(t) = H_0(t) + Yf(t)$.
In an interaction picture that captures all influences but the weak
additional driving, the Liouville-von Neumann equation reads $\dot\rho =
-i[Y(t),\rho]f(t)$.  Its integrated form
provides the first-order solution
\begin{equation}
\rho(t) = \rho_0(t) -i\int_{-\infty}^t dt'\, [Y(t'),\rho_0(t')] f(t') \,,
\end{equation}
such that the expectation value of an operator $Z$ reads
\begin{equation}
\label{app:kubo}
\langle Z_t\rangle = \langle Z_t\rangle^{(0)} +\int dt'\, \chi(t,t') f(t')
\end{equation}
with the susceptibility
\begin{equation}
\chi(t,t') = -i\langle[Z(t),Y(t')]\rangle_{t'}^{(0)} \theta(t-t') \,.
\end{equation}
Formally, this is the usual Kubo formula, but with the equilibrium density
operator replaced by the non-equilibrium $\rho_0(t')$ which may depend on
the dynamics of the strongly driven qubit as well as on its initial state.
The latter is already the case for the usual dispersive readout
\cite{Blais2004a} by which one determines whether the qubit is initially in
the ground state or in the excited state.  Notice that the susceptibility
$\chi(t,t')$ depends explicitly on both times such that the integral in
Eq.~\eqref{app:kubo} generally is not a mere convolution.

\section{Susceptibility of a periodically driven system}
\label{app:susceptibility}

If the Hamiltonian $H_0(t)$ is $T$-periodic, two-time expectation values
such as the susceptibility in the long-time limit are invariant under a
shift of both times, $\chi(t,t') = \chi(t+T,t'+T)$.  Then introducing the
time difference $\tau=t-t'$ yields $\chi(t,t-\tau) = \chi(t+T,t+T-\tau)$.
Consequently, the frequency representation of $\chi(t,t-\tau)$ can be
written as a Fourier series in $t$, while the $\tau$-dependence still
requires a Fourier integral \cite{Kohler2005a}.  Therefore,
\begin{equation}
\label{app:chi}
\chi(t,t-\tau) = \sum_k\int \frac{d\omega}{2\pi} e^{-ik\Omega t}
e^{-i\omega\tau} \chi^{(k)}(\omega)
\end{equation}
with $\Omega=2\pi/T$.  Inserting this expression into Eq.~\eqref{app:kubo}
results in the linear response formula in Fourier space,
\begin{equation}
Z_\omega  = Z_\omega^{(0)}
+ \sum_k\chi^{(k)}(\omega-k\Omega) f_{\omega-k\Omega} \,.
\end{equation}
It reveals that the reaction of a periodically driven quantum system to a
weak probe $f(t)$ is characterized by frequency mixing with sidebands
separated by the frequency $\Omega$ of the strong driving.

\subsection{Floquet theory}

The explicit computation of the susceptibility may still be a formidable
task.  Here, we perform it within the Floquet-Markov approach of
Refs.~\cite{Kohler1997a, Blattmann2015a}. It starts from a system-bath
model \cite{Leggett1987a, Hanggi1990a} in which the qubit couples to an
ensemble of harmonic oscillators described by the Hamiltonian
$H_\text{bath} = \sum_\nu \hbar\omega_\nu b_\nu^\dagger b_\nu$, where
$b_\nu$ is the usual bosonic annihilation operator of an oscillator
with frequency $\omega_\nu$.  The bath oscillators couple to a qubit operator
$X$ according to the Hamiltonian
\begin{equation}
\label{bathcoupling}
H_\text{qb-bath} = X \sum_\nu \lambda_\nu (b_\nu^\dagger+b_\nu) \,.
\end{equation}
The influence of the bath can be captured by its spectral density
$J(\omega) = \pi\sum_\nu|\lambda_\nu|^2 \delta(\omega-\omega_\nu) \equiv
J(\omega) = \pi\alpha\omega/2$ with the dimensionless coupling strength
$\alpha$ \cite{Leggett1987a, Hanggi1990a}.
Within second-order perturbation theory one finds for the reduced qubit
density operator the Bloch-Redfield master equation
\cite{Redfield1957a, Blum1996a}
\begin{equation}
\label{app:BR}
\begin{split}
\dot\rho ={}& -i[H_\text{qb}(t),\rho]
\\& -\int_0^\infty d\tau\big\langle [H_\text{qb-bath},[\tilde
H_\text{qb-bath}(t,t-\tau),\rho]\big\rangle_\text{bath,eq} .
\end{split}
\end{equation}
For rather weak dissipation, $\rho(t)$ eventually becomes diagonal in the
(time-dependent) basis of the Floquet states,
\begin{equation}
\label{app:rho}
\rho_0(t) = \sum_\alpha p_\alpha
|\phi_\alpha(t)\rangle\langle\phi_\alpha(t)| \,,
\end{equation}
with the occupation probabilities $p_\alpha$.
For this diagonal ansatz, the first term in Eq.~\eqref{app:BR} vanishes
such that one obtains the Pauli-type master equation
\begin{equation}
\label{master}
\dot p_\alpha = \sum_\beta\big( W_{\alpha\leftarrow\beta}p_\beta -
W_{\beta\leftarrow\alpha} p_\alpha \big)
\end{equation}
with the generalized golden-rule rates
\begin{equation}
W_{\beta\leftarrow\alpha}
= 2\sum_k \big|X_{\beta\alpha,k}\big|^2
N(\epsilon_\beta-\epsilon_\alpha-k\Omega) ,
\end{equation}
and the Fourier components of the transitions matrix elements,
\begin{equation}
\label{app:Xabk}
X_{\beta\alpha,k} = \int_0^T \frac{dt}{T}\, e^{ik\Omega t}
\langle\phi_\beta(t)|X|\phi_\alpha(t)\rangle \,.
\end{equation}
The function $N(\omega) = J(\omega)n_\text{th}(\omega)$ contains the
bosonic thermal occupation number $n_\text{th}(\omega) =
[e^{\hbar\omega/k_BT}-1]^{-1}$.  In order to arrive at this concise form,
we have defined the spectral density for negative $\omega$ as $J(\omega) =
-J(-\omega)$, while the Bose function has been extended by analytic
continuation.

Having obtained the time-independent long-time solutions for the
probabilities $p_\alpha$, we can straightforwardly evaluate the
susceptibility \eqref{app:chi} which for $Y=Z$ results in the Fourier
components
\begin{equation}
\label{app:chiFloquet}
\chi^{(k)}(\omega) = \sum_{\alpha,\beta,k'} \frac{(p_\alpha-p_\beta)
Z^*_{\beta\alpha,k'-k} Z_{\beta\alpha,k'}}
{\omega+\epsilon_\alpha-\epsilon_\beta +k'\Omega +i\gamma/2} \,,
\end{equation}
where the matrix elements $Z_{\beta\alpha,k}$ are defined according to
the $X_{\beta\alpha,k}$ in Eq.~\eqref{app:Xabk}.

\subsection{Limit of adiabatic following}

\begin{figure}[tb]
\centerline{\includegraphics{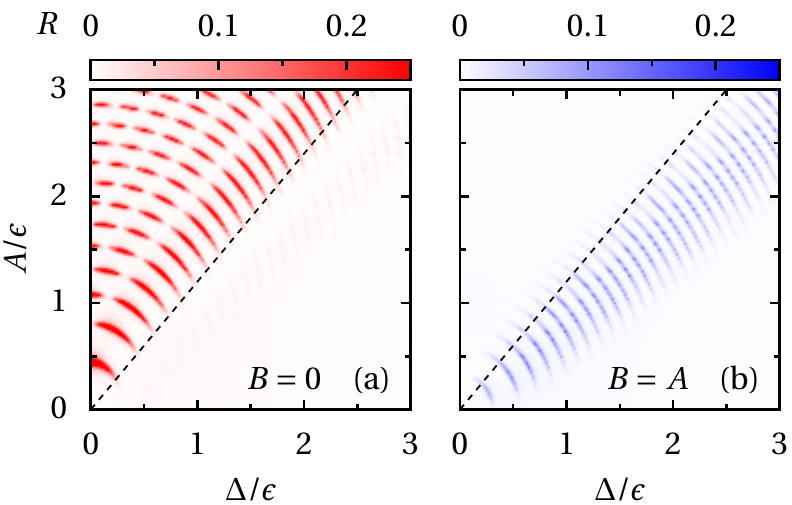}}
\caption{Cavity reflection for the alternative Hamiltonian
$H_\text{qb}'(t)$ for one-parameter driving with $B=0$
(a) and two-parameter driving with $B=A$ (b).  Here the detuning $\epsilon$
is kept constant, while the tunneling $\Delta$ is varied.
All other parameters are as in Fig.~\ref{fig:patterns}.
The dashed lines mark the values used in Fig.~\ref{fig:berryphase}(b).
}
\label{fig:patterns2}
\end{figure}

It is instructive to relate the susceptibility~\eqref{app:chiFloquet} to
the corresponding expression in the adiabatic limit
based on the adiabatic solutions of the Schr\"odinger equation,
\begin{equation}
\label{app:psiad}
|\psi_\alpha(t)\rangle = e^{i\varphi_\alpha(t)}|u_\alpha(t)\rangle \,,
\end{equation}
where the phase of each adiabatic eigenstate $|u_\alpha(t)\rangle$ is
determined by
$\dot\varphi_\alpha = \langle u_\alpha(t)| i\partial_t|u_\alpha(t)\rangle
-E_\alpha(t)$.  Integrating this equation of motion over one driving period
yields
\begin{equation}
\label{app:phiT}
\varphi_\alpha(t+T)-\varphi_\alpha(t)
= \gamma_\alpha^\text{ad} -\bar E_\alpha T
\end{equation}
with the adiabatic phase $\gamma_\alpha^\text{ad}$ and the dynamical phase
$-\bar E_\alpha T$ determined by the mean energy $\bar E_\alpha$.  To make
use of the time-periodicity that allows one to bring the susceptibility to
the form of Eq.~\eqref{app:chi}, we write the total phase as the sum of a
linearly growing contribution and a $T$-periodic part
$\tilde\varphi_\alpha(t)$,
\begin{equation}
\varphi_\alpha(t) = (\gamma_\alpha^\text{ad}/T-\bar E_\alpha)t
+ \tilde\varphi_\alpha(t) \,,
\end{equation}
which can be understood as definition of $\tilde\varphi_\alpha(t)$ whose
$T$-periodicity follows from Eq.~\eqref{app:phiT}.  The
term in brackets reminds one to the quasienergy expressed by the
geometric phase and the mean energy of a Floquet state, as is discussed in
the main text.

To put this correspondence on a solid ground, we use the fact that within
the adiabatic approximation, $|\psi_\alpha(t)\rangle$ solves the
time-dependent Schr\"odinger equation.  Then the state
\begin{equation}
|\phi_\alpha^\text{ad}(t)\rangle
= e^{i\tilde\varphi_\alpha(t)}|u_\alpha(t)\rangle
\end{equation}
on the one hand obeys the Floquet equation and on the other hand is
$T$-periodic.  Therefore it is a Floquet state with a quasienergy
determined by the phase acquired during one driving period,
$\epsilon^\text{ad}_\alpha = \bar E_\alpha -\gamma_\alpha^\text{ad}/T$.

With the Floquet form of the adiabatic eigenstates at hand, it is
straightforward to evaluate the susceptibility.  We assume that
for slow driving the system follows the adiabatic ground state, such that
the populations read $p_\alpha = \delta_{\alpha,0}$.  Neglecting in
Eq.~\eqref{app:chiFloquet} the counter-rotating contributions, i.e., those
with $-p_\beta$ and $\beta\neq\alpha$, we find for the adiabatically driven
quantum system the susceptibility
\begin{equation}
\chi^{(k)}(\omega) = \sum_{\beta\neq0,k'}
\frac{Z^*_{\beta 0,k'-k}Z_{\beta 0,k'}}
{\omega+\epsilon^\text{ad}_0-\epsilon^\text{ad}_\beta+k'\Omega+i\gamma/2} \,.
\end{equation}
Numerical tests demonstrate that for sufficiently low driving frequencies,
the Floquet theory and the adiabatic theory indeed yield the same cavity
reflection.  This agreement also verifies the scaling of the resonance
condition as a function of $\Omega$ upon which the measurement scheme for
the adiabatic phase is based.

\section{LZSM pattern for the alternative Hamiltonian}
\label{app:patterns}

Figure~\ref{fig:patterns2} shows the LZSM patterns for the
alternative Hamiltonian
\begin{equation}
H_\text{qb}'(t) = \frac{\Delta}{2}\sigma_x + \frac{\epsilon}{2}\sigma_z
+A\sigma_x\cos(\Omega t) -B\sigma_y\sin(\Omega t) \,,
\end{equation}
with the couplings to the bath and to the cavity specified as
$X=Z=\sigma_z$.  For both one-parameter driving (panel a) and two-parameter
driving (panel b), the most pronounced resonances lie slightly above the
bisecting line.  For $A=1.2\Delta$, the readout signal turns out to be
sufficiently strong for recovering the adiabatic phases, which motivates
the choice of the amplitudes in Fig.~\ref{fig:berryphase}(b) of the main
text.  As in the case of the Hamiltonian $H_\text{qb}(t)$, for
two-parameter driving with equal amplitudes, the inner structure of the
pattern vanishes, while only the outermost resonances remain.

%

\end{document}